\documentclass[referee]{aa} 
\usepackage{graphicx}
\begin{document}
\title{ROTATION VELOCITIES OF HOT HORIZONTAL BRANCH STARS IN THE GLOBULAR CLUSTERS 
NGC~1904, NGC~2808, NGC~6093, AND NGC~7078: THE DATABASE
\thanks{Based on observations with the ESO {\it Very Large Telescope + UVES}, ~at the 
Paranal Observatory, Chile}}

\author{A. Recio-Blanco\inst{1}, G. Piotto\inst{1}, A. Aparicio\inst{2}, A. Renzini\inst{3}}

\offprints{A. Recio-Blanco}

\institute {Dipartimento di Astronomia, Universit\`a di  Padova,   
Vicolo  dell'Osservatorio  2, I-35122     Padova, Italy\\
\email{recio,piotto@pd.astro.it }
\and 
Instituto de Astrofisica de Canarias, Via Lactea s/n,
382002 La Laguna Tenerife, Spain\\
\email{aaj@ll.iac.es}
\and  ESO, Karl-Schwarzschild-Str. 2, D-85748 Garching bei 
M$\ddot{u}$nchen, Germany\\  
\email{arenzini@eso.org}
}

\date{Received May .., 2003; accepted .., 2003}

\abstract{

We present radial and rotation ($v$sin$i$) velocity measurements, 
from UVES+VLT high resolution spectra of 61 stars in the blue 
horizontal branches (HB) of the Galactic globular clusters NGC~1904 (M79), 
NGC~2808, NGC~6093 (M80), and NGC~7078 (M15). 

The data reduction and the velocity determination, based on 
cross-correlation techniques, are discussed in detail.  Most of this 
database has been used by Recio-Blanco et al. (2002) in their analysis 
of the rotation velocity properties of blue HB stars.  Here we present 
additional data for NGC~2808. We confirm the results of the previous 
paper, in particular, a possible link between the HB jump and the 
abrupt change in the rotational velocity distribution around T$_{\rm 
eff}$ $\sim$ 11,500 K.

}

\authorrunning{Recio-Blanco et al.}

\titlerunning{Rotation of hot HB globular cluster stars.}

\maketitle

\section{Introduction}

The high  resolution  spectroscopic  capabilities of  the  Ultraviolet
Visual Echelle Spectrograph (UVES), combined with the collecting power
of the Very Large Telescope (VLT), offer an exceptional opportunity to
study the nature of the  hot stars populating the anomalously extended
blue horizontal branches (HB) in  Galactic globular clusters (GC). The
horizontal  branches with blue tails  (BT{\bf, here considered to
host stars hotter than 12000 K}) probably represent the most
extreme of   the mixed  bag of  HB  anomalies  lumped  under  the term
``second-parameter   problem'' (cf. Catelan et    al. 1998, Piotto  et
al. 1999 for a more detailed discussion).
We already know that BTs host  He-core H-shell burning stars which have
lost up to a factor of two more mass during the Red Giant Branch (RGB)
ascent, than other HB stars in the same cluster (D'Cruz et al., 1996).
However, we still lack of an explanation for such a sizable mass loss.
Several  mechanisms, regulating either the  core mass or the amount of
mass loss that a star experiences on the  upper RGB have been proposed
as possible second parameter candidates. Among them, stellar rotation,
which would delay the  helium  core flash in a   red giant, was   also
suggested  (Mengel \&  Gross,  1976; Renzini,  1977;  Peterson et al.\ 
1995).    A faster rotating  star  would therefore increase its helium
core mass and would  experience a higher   mass loss rate in the  RGB,
leading to bluer and brighter HB star.

The first investigations on the rotation velocity of {\bf blue horizontal branch} 
stars were done by Peterson and collaborators (Peterson 1983, 1985a, 
1985b, Peterson et al.\ 1995). Some of the blue HB stars in the BT cluster M13 
were found to be rotating as fast as 40 km/s. On the contrary,
clusters with cooler HB morphologies, as M3 and NGC~288, 
showed {\bf only} values of $v$sin$i$ slower than about 20 km/s.
More recently, Behr et al.\ (2000a, 
B00a) have suggested the existence of a discontinuity in stellar 
rotation velocity across one region underpopulated of stars (gap), at 
T$_{\rm eff}$ $\simeq$ 11.000 K, in the {\bf HB} of M13. Bluewards of the 
gap, all the stars showed modest rotations ($v$sin$i < 10$ km 
s$^{-1}$), while, to the red side of the gap, several rapidly rotating 
stars were found.   
A similar discontinuity was also detected in M15 (Behr et al. 2000b, 
B00b). 
 
UVES+VLT allows {\bf us} to measure the rotational velocity of blue HB stars as 
faint as $V=18$ with an accuracy of a few km/s.  For this reason, and 
prompted by the previous results of B00a, we started a program to 
investigate how stellar rotation changes as a function of the position 
along the HB, and from cluster to cluster. We observed at least ten stars 
per cluster, to statistically avoid the problem of the projection angle
(sin$i$) in the determination of stellar rotation. At the present time, we have analyzed 61 stars 
in 4 Galactic GCs: NGC~1904 (M79), NGC~2808, NGC~6093 (M80) and NGC~7078 
(M15). 
All the target clusters have a blue tail HB morphology, although their horizontal 
branches  have different temperature extensions (see their colour-magnitude diagrams in Fig. 1). The 
HB of M79 reaches T$_{eff}$ $\sim$ 23,000 K, {\bf both M80 and M15 BTs reach a
T$_{eff}$ around 30,000 K.}, while  the HB in NGC~2808 goes  up 
to T$_{eff}$   $\sim$  37,000 K.  They also span a range of metallicity,
total luminosity, concentration,... as summarized in Table 1.

The target stars have a temperature in the range $8,000 K 
< T_{\rm eff} < 28,000 K$, distributed on the two sides of the 
Str$\ddot{\rm o}$mgren ({\it u, uy}) luminosity ``jump'' at $T_{\rm 
eff} \simeq 11,500 K$ (Grundahl et al. 1999, G99). Most of the 
database we are going to present in this paper has already been used 
by Recio-Blanco et al. (2002,R02), with a number of interesting results.  
For the first time, we revealed the presence in NGC~1904 of a considerably large
number  of fast 
($v$sin$i > 20$ km/s) HB rotators, and confirmed the fast rotators 
already detected in M15 by B00b. In both cases, fast rotators were 
confined to the cool end of the blue HB, as in M13.     
In this paper, we add 5 more stars in NGC~2808, improving statistics 
on the cool side of the G99 jump for this cluster. We also discuss in 
detail the reduction and analysis techniques (Sections 2, 3 and 
4). In addition, we present the position and the radial and projected 
rotational velocities for all the stars so far observed (Section 4). A brief
discussion on the results (Section 5) is also included.

\section{Observations and Data Reductions}
\label{s_obs}

We selected our targets from three different photometric data sets: 
the F439W and F555W HST-WFPC2 
photometry by Piotto et al. (2002), the ground-based Johnson V, I photometry by 
Rosenberg at al. (2000), and the Johnson V,B,U data from Bedin et 
al. (2000). In addition, we have also identified the NGC~1904 targets 
in the Str$\ddot{\rm o}$mgren {\it u, y} photometry by G99. 
 
The observations were carried out with the UVES echelle spectrograph, 
mounted on the Unit 2 (Kueyen) of the VLT, and the 2K x 4K, 15 $\mu$m 
pixel size blue CCD, with a readout noise of 3.90 e$^{-}$ and a conversion 
factor of 2.04 e$^{-}$/ADU.
The  UVES  blue    arm, with a  spectral  coverage    in the 3730-5000
$\stackrel {\circ}  {\rm A}$ range,  was used.  Combined with  a  slit
width of 1.0 arcsec, we   achieved  a resolution of R$\sim$   40$~$000
($\delta \lambda \sim 0.1 \stackrel {\circ}  {\rm A}$, $\delta v \sim$
7.5 km/s, see the UVES   user manual by  Kaufer  et al.\ 2003).   
We decided to use only the UVES blue arm because there were  very
few lines in the red arm spectral range useful for the measurement of
stellar rotation, specially for the hottest stars.
The spectra were obtained during 3 observing runs: July 30--August 2 2000,
January 19--23 2001, and March 1--7 2002 (service mode).
The exposure times range from 800 s (for V $\sim$ 16) to 8500 s (for V
$\sim$ 18.5).  We generally  limited individual exposure times to 1500
or 2000 seconds, to minimize cosmic ray accumulation, and then coadded
2 to 4 frames per  star. The typical signal to  noise ratios,{\bf measured
with the task {\it splot} within IRAF}, are about
10 to 15 per pixel, but in some cases they {\bf achieve a value of} S/N $\sim$ 20-30
(see Tables 2 and 3 for a detailed information).
In general, all the targets lie in the low-crowding outskirts 
of the parent cluster, to avoid contamination from other stars. {\bf The
isolation of the targets was checked directly, usingthe corresponding avaliable images
of each cluster.}
During each observing run, we also collected high S/N spectra of a set of 
field rotational velocity standards (Peterson 1983), with spectral 
types close to those of our program stars: HD~74721 ($v$sin$i$ $<$ 6 km/s), 
HD~130095 ($v$sin$i$ $<$ 6 km/s), HD~213468 ($v$sin$i$ $<$ 10 km/s),
HD~117880 ($v$sin$i$=12$\pm$3 km/s), HD~19445 ($v$sin$i$=13$\pm$3 
km/s), and HD~109995 ($v$sin$i$=27$\pm$3 km/s). 
Peterson (1983) used the cross-correlation technique to find {\bf these} values,
after a calibration with an artificially broadened spectrum
of a non rotating HB field star, which was also chosen as the template 
(see Peterson 1983, Section IVe). The resolution of her observations  was 8.5$\pm$0.5 km/s.
 
The spectra were extracted manually, trying to achieve the highest possible
signal to noise. Standard IRAF procedures were used. 
For the wavelength calibration,   ThAr spectra were obtained  at  the
beginning and the  end of each night. Thanks  to its technical design,
UVES  is an  instrument  with a  high stability  and  repeatability of
calibrations both  over  short   and  extended periods of  time.    In
addition,   the spectrograph  is   placed  at   the  {\bf Nasmyth} platform
minimizing flexure problems.  Changes   within a night of the atmospheric
pressure and the temperature inside the enclosure were always smaller than
1 hPa  and   1.5 $\deg$C respectively,   so  the corresponding induced
errors are $<$ 50 m/s and $<$ 250 m/s (see UVES  user manual). This is
empirically confirmed by the ThAr spectra  taken at the beginning and
the end of each night. 
The seeing disk values oscillated in the range $\sim$ 0.5 arcsec to 
$\sim$ 1.4 arcsec. As a consequence,  the possibility that apparent 
stellar radial velocity could depend a little on the star's position 
within the slit has to be taken into account, when the seeing profile was
narrower than the slit. On the other hand, the wind speed was always
$\leq$ 10 m/s so possible errors derived from guiding drift are always
very small (a few hundreds of meters per second, see VLT documentation).
{\bf Moreover, empirical tests on radial velocity
performed on our data (cf. Section 3) confirm that there
are not large v$_{rad}$ errors due to guiding drift.}

We  fitted 4$^{th}$ order polynomials  to the  dispersion relations of
the  ThAr calibration  spectra,  which  resulted  in residuals  $\leq$
3$\cdot$10$^{- 4}$ $\stackrel    {\circ} {\rm A}$. 
A mean of 4 to 5 lines per  order were used for
the polynomial wavelength   solution,   covering properly all  the   31
orders.Each  spectrum  was
divided into sections of about  40 $\stackrel {\circ} {\rm A}$, which,
avoiding  hydrogen lines, in general   coincided with the echelle UVES
orders. 
We also avoided  the broad CaII K line at 3933 $\stackrel {\circ}  {\rm A}$.
Finally, each  spectrum was  normalized   using a manual polynomial
fitting.

\section{Cross-correlation analysis}
\label{cc}

We determined the radial velocity (v$_{rad}$) and the projected 
rotational velocity ($v$sin$i$) for each of our program stars using 
the cross-correlation technique described by Tonry \& Davis 
(1979,TD79). This method is well {\bf suited} to measure v$_{rad}$ and 
rotational broadening in low signal-to-noise spectra (Dubath et al.\ 
1990).  The analysis procedure computes (in the Fourier domain) the 
cross-correlation function (CCF) of the object spectrum versus that of 
a template, fits a Gaussian to the highest peak, and finds the radial 
velocity and the line broadening from the peak's central position 
($\delta$) and width ($\sigma$).  If both spectra are very 
narrow-lined, there is little tolerance for radial velocity 
displacements, and the peak falls off rapidly; if one or the other is 
somewhat broad-lined, the falloff is gradual, and the correlation peak 
is broader. The cross-correlation procedure was performed using the task {\it fxcor} 
within IRAF. For the Gaussian fit of the CCF peak the parameters used 
were the center position, amplitude and width. {\bf Separate CCFs were
computed for each spectral order}.
The heliocentric correction for each spectrum was also computed 
on the basis of its observational coordinates and time.  We used the 
slow rotating ($v$sin$i <$ 6 km/s) blue HB field stars HD~74721 
(v$_{rad}$=30.7 km/s $\pm$ 0.6 km/s) and HD~130095 (v$_{rad}$=66.0 km/s $\pm$ 0.7 km/s) as templates 
for the cross-correlation. Those CCFs for which the peak's central 
position did not agree with the correct stellar radial velocity, 
relative to the template, were rejected. In general, we considered only
those CCFs giving an object v$_{rad}$  in the range v$_{rad_{GC}} \pm$ 50 km/s,
where v$_{rad_{GC}}$ is the radial velocity of the corresponding globular cluster,
as taken from Harris (1996) compilation.  In this way, we avoided the errors 
due to line mismatch or to the lack of spectral features, especially 
for the hottest stars and the reddest sections of the spectra.  A mean 
of 15 different CCFs could be used for each program star, each of them 
providing a value of $\delta$ and $\sigma$. In addition, using 
$\delta$ as the center position, the antisymmetric part of the 
correlation function is calculated, and its rms ($\sigma$$_{a}$) is 
derived. This allows the calculation of the velocity error 
($\epsilon$) via the ``r'' parameter, as described by TD79: 
\begin{center} 
r$ = h / (\sqrt{2}\sigma_{a})$ 
 
$\epsilon = N / (8B (1 + r)) $ 
\end{center} 
 
where h is the height of the cross-correlation peak, N is the number of bins in which the spectrum is 
mapped (2048 in our case) and B is the highest wave number where the Fourier transform of the CCF has 
appreciable amplitude (e.g. the half-maximum point). 
 
To calculate the radial velocity of each program star, we computed the 
mean ($\delta_{mean}$) of the $\delta$ values of all the good CCFs 
involving that star, weighted to the inverse of their errors 
(1/$\epsilon$).  The final v$_{rad_{o}}$ of the object star is: 
 
\begin{center} 
v$_{rad_{o}} = v_{rad_{t}} + \delta_{mean}$ 
\end{center} 
 
where v$_{rad_{t}}$ is the {\bf most recent} radial velocity of the template taken {\bf from} the literature (Kinman et al 2000). 
Hence, the error:  
\begin{center} 
$\epsilon$(v$_{rad_{o}}$) = $\sqrt{\epsilon^{2}(v_{rad_{t}}) + \epsilon^{2}(\delta_{mean})}$    
\end{center} 
 
where $\epsilon(v_{rad_{t}})$ is taken from Kinman et al.\ (2000) and 
$\epsilon(\delta_{mean})$ is the rms scatter of the individual $\delta$ values.
In order to take into account possible systematic errors on the v$_{rad}$ measurements
induced by the adopted values of v$_{rad_{t}}$, we have tested the internal consistency
among the field stars used. Cross-correlations among {\bf all the spectra  of the 
stars HD~74721 and HD~130095 observed (seven spectra for each star)} confirm an 
agreement with Kinman et al.\ (2000) inside the errors and
a consistency of $<$ 0.6 km/s among the standards.

Similarly, after the appropriate calibration, we obtained the $v$sin$i$ 
for each target from the value of $\sigma$, as explained in the 
following section.  Figure 2 shows an example of the CCFs obtained and 
fitted. The target star is n2808-2333, and the template is HD~74721. The 
corresponding normalized spectra sections are presented in Figure 3. 
The background level is always fixed at the zero value, on the complete range of the CCF 
to avoid the sensitivity of the results to  details  of  the  shape  of  the 
correlation  function.

\section{Projected rotational velocity measurements}

The width ($\sigma$) of the CCF of a star results from several 
broadening mechanisms which depend on, for instance, gravity, 
turbulence, magnetic fields, effective temperature, metallicity, 
rotation.  In addition, the instrumental profile also contributes to 
the broadening of the spectrum and therefore to the CCF. Thus, in 
order to correctly measure the rotational contribution to the width of 
the CCF, we must model the contributions of the broadening mechanisms 
other than rotation, that is, calibrate the $\sigma$ - $v$sin$i$ 
relation for the instrument and the stars we are working with.  In the 
case of a Gaussian fit of the CCF peak, the rotational broadening will 
correspond to a quadratic broadening of the CCF: 
\begin{center} 
$\sigma^{2} = \sigma^{2}_{rot} + \sigma^{2}_{o} \Rightarrow \sigma^{2}_{rot} = \sigma^{2} - \sigma^{2}_{o}$ 
\end{center} 
where $\sigma$ is the width of the CCF, $\sigma_{rot}$ is the rotational broadening, and 
$\sigma_{o}$ gives the non rotational contributions to the width of the CCF 
from both the object star and from the template, which we suppose to be equal (same
spectral type and same instrumental broadening).
Therefore, the projected rotational 
velocity $v$sin$i$ is given by (Benz \& Mayor 1984): 
\begin{center} 
$v$sin$i$ = $ A \cdot \sqrt{\sigma^{2} - \sigma^{2}_{o}} = A \cdot \sigma_{rot}$    (1) 
\end{center} 
where A is a constant coupling the differential broadening of the CCF to the 
difference in $v$sin$i$ between the template and the object. 
The value of A in the previous equation was found  
by fitting a straight line to the relation, $v$sin$i$ versus 
$\sigma_{rot}$, for all the field standards of rotation, assuming
3 km/s for the rotation of the templates. The slope
gives A. The best fit of the points, $<$ A $>$ = 1.8 $\pm$ 0.3, was adopted 
(see Figure 4). 
The previous relation was  observed to be linear in our range of interest,
and even for faster rotation values, by Lucatello \& Gratton (2003, see their
Figure 2). This is also confirmed by the very low residuals ($<$ 0.1) that we find
by fitting the cross-correlation peaks with a Gaussian profile.

As explained  by Melo et al (2001),  there is also a smooth dependence
on  $\sigma_{o}$ on the stellar  colour and luminosity class. However,
the  variation  of $\sigma_{o}$ within  the range  of  our program and
template stars is expected to be  quite small compared to the accuracy
of our $\sigma$ measurements ($\sim$ 3-6 km/s),  due to the relatively
small  number of lines  and the low S/N for  our spectra. Moreover, we
have an error of 3 km/s in the $v$sin$i$ values of the templates.  For
this reason,  we decided  to  estimate the  value of  $\sigma_{o}$  by
correlating one template with  the other, for each  one of  the orders
used to measure the $v$sin$i$ of the target stars.
In other  words,  $\sigma^2_{o}$  is equal  to  $\tau^2_1$+$\tau^2_2$,
where  $\tau_1$ and $\tau_2$ are  the line widths  of the templates, in
this case  HD~74721 and HD~130095.  The value  of $\sigma_{o}$ changes
very {\bf little} with the echelle order used or the night of observation, and it has a
mean value of 13.3 $\pm$ 0.4 km/s.
In order  to  test the  consistency of  the previous approximations, we
have   estimated   the values  of   $\tau_1$   and $\tau_2$  from  the
autocorrelations  of  HD~74721 and    HD~130095.  As in  the  case  of
$\sigma_{o}$, $\tau_1$ and $\tau_2$ are  quite constant with the order
and the night   of observation  and  the mean  values  determined  are
$\tau_1 =$ 12.8 $\pm$ 0.3 km/s and $\tau_2  =$ 13.3 $\pm$ 0.6 km/s. To
avoid an   underestimation   on the   error,   we have cross-correlated
different spectra of the same template,  instead of autocorrelating an
spectrum with itself. 
These results  suggest that the $v$sin$i$ value  is quite the same for
both  templates,   specially   considering their  similar  atmospheric
parameters (Kinman  et al.\  2000).  Nevertheless, in  the case of the
hotter targets in our  sample,  we could be  slightly  overestimating
their rotational rates if their $\sigma_{o}$ were somewhat larger than
the one  of our templates.  However,  for stars with temperatures  $>$
11500 K, the absence  of microturbulence and surface convection could,
at least partially, compensate the previous effect.   In any case, the
overall  effect  is expected   to  be rather  small  compare to the
accuracy of our measurements.

In order  to  derive the final  $v$sin$i$ for   each program  star, we
calculated the mean  of the $v$sin$i$  values given by the  individual
$\sigma$ determinations  for all   the   good  CCFs (15 on    average)
involving that   star. We    used  as  weight   the  height    of  the
cross-correlation peak.

The error of each $v$sin$i$ measurement can be obtained by differentiating Eq. 1: 
\begin{center} 
$|\Delta(v$sin$i)|_{mes} = A^{2}  \sqrt{(\epsilon  \sigma)^{2} + (\epsilon_{o}  \sigma_{o})^{2}} / v$sin$i$ 
\end{center} 
 
where $\epsilon$ and  $\epsilon_{o}$  are the velocity errors  for the
CCF and the template to template CCF respectively. To  this we have to
add the systematic errors due to the uncertainty in the templates and
the standards, which also affect the calibration curve. A first systematic
error, $|\Delta(v$sin$i)|_{t}$, has been evaluated by re-calculating 
the $v$sin$i$ of each object star assuming a $v$sin$i$ for the templates
of 0 km/s and 6 km/s. A second systematic, $|\Delta(v$sin$i)|_{A}$, 
error was derived by re-calculating the $v$sin$i$ of the targets with A 
set to its $+\sigma$ and $-\sigma$ values.
Thus, the  total error on $v$sin$i$ is:

\begin{center} 
$|\Delta(v$sin$i)|_{tot} = \sqrt{|\Delta(vsini)|^{2}_{mes} + |\Delta(vsini)|^{2}_{t} + |\Delta(vsini)|^{2}_{A}}$ 
\end{center} 

\section{Results}

The final results for a dataset of 61 stars in the clusters NGC~1904 
(M79), NGC~2808, NGC~6093 (M80) and NGC~7078 (M15) are presented in 
Tables 4 and 5. For each target star (Column 1), the radial velocity (Column 2), and the 
projected rotational velocity (Column 3) are given. The three sources of error
(random error, error due to the templates uncertainty, and error due to the calibration
curve) are presented in Columns 4, 5 and 6. The total error, resulting from the
quadratic addition of random and systematic errors is given in Column 7. 
The M79 Str$\ddot{\rm o}$mgren photometric data from G99 have been kindly 
provided by F. Grundahl. The NGC~2808 {\it U, U-B} photometry is from 
Bedin et al. (2000). The M80 and M15 photometry comes from Piotto et 
al. (2002), except for 5 stars in M80 and 3 stars in M15, taken from 
Rosenberg et al. (2000). In addition, the effective temperature has 
been derived by comparing the Cassisi et al. (1999) models with the 
corresponding colour-magnitude diagrams. 
We tested the consistency of
the derived T$_{eff}$ values from different photometric colours comparing the
Str$\ddot{\rm o}$mgren photometry  of cluster M79 with a Johnson photometry
by Momany et al.\ (2003, private communication). The temperature differences 
were always $\leq$ 1000 K.

 The stars m15-b130 and m15-b218 are in common with B00b. 
Our $v$sin$i$ measurements and B00b values for this two stars are in 
agreement within the errors (5 $\pm$ 3 vs. 5.07 $\pm$ 0.24 and 13 
$\pm$ 3 vs. 14.88 $\pm$ 0.69), which reveals the consistency between both
observations and the two independent methods applied for 
the rotational velocity measurement (B00 used a profile fitting method). 
 
Figure 4 shows the complete set of projected rotational velocity data 
from 5 BT Galactic GCs: M79, NGC~1904, M80, M15 and M13. Besides the 
data from this paper (circles), Fig. 4 includes the B00b data for M15 
(open triangles), plus the B00a (full triangles) and {\bf Peterson et al.\ (1995,} full 
squares) measurements for M13. 
 
As already pointed out in R02, all the stars 
hotter than T$_{\rm eff}$ $\sim$ 11,500 K 
have $v$sin$i\le$ 12 
km/s, indicating that the bulk of these stars must be intrinsically 
slow rotators. At T$_{\rm eff}$ $\sim$ 11,500 K (which is the 
temperature of the $u$-jump of G99) there is an abrupt change in the 
rotational velocity distribution.  Among the cooler stars (T$_{\rm 
eff}$ $<$ 11,500 K), there is a range of rotation rates, with a group 
of stars rotating at $\sim$ 15 km/s or less, and a fast rotating group 
at $\sim$ 30 km/s.  A few stars reach a projected rotation velocity of 
about 40km/s.  As noticed by R02, the fraction 
of fast rotators is different in different clusters (cf. Fig. 1 and 2 
in their paper), with fast rotators relatively more abundant in M13 
and M79 than in the other clusters of the present sample. In M13 and 
M79, at least half of the stars cooler than T$_{\rm eff}$=11,500 K are fast 
rotators, while only 3 out of 22 stars rotate faster than 15km/s in 
M15.  At the present time, no fast rotators have been identified in NGC~
2808 and M80.  Figure 5 shows the projected rotational velocities for 
the 16 target stars in NGC~2808, with the new stars presented in this 
paper added.  All the newly observed stars are cooler than T$_{\rm 
eff}=$11,500 K.  Again, none of them has a projected rotation velocity 
higher than 13 km/s. However, on the cool side of the gap, we have rotation velocities for only 9 
stars in NGC~2808 and 7 stars in M80.  If 
the fraction of fast rotators in these two clusters is as low as in 
M15 (or lower), we might have missed them because of the small number 
of observed targets. The rotation rate distribution, and its variation
from cluster to cluster must be studied. The advent of the multifiber spectrograph
FLAMES at VLT will provide a unique opportunity to carry out this project. 
In particular, we plan to continue our study with new observations
to increase statistics of rotational velocity measurements.

Figure 4 clearly suggests a link between the photometric G99 jump and the
discontinuity in the distribution of the HB stellar rotation rates.
Radiative levitation of metals has been invoked by G99 to explain the jump.
Besides, the metal abundance anomalies (enhancements of metals,
underabundance of He) found by Behr et al (1999) and B00b constitute an
empirical evidence that radiative levitation and diffusion are
effectively at work in the envelope of HB stars with $T_{\rm eff}\ge
11,500$K. As a consequence, R02 argue that  the
absence of fast rotators at $T_{\rm eff}\ge 11,500$K might be due to an increase of 
angular momentum removal caused by enhanced mass loss in the more 
metallic atmospheres of the stars hotter than the jump.
An enhanced mass loss in higher{\bf metallicity} atmospheres has been very
recently calculated by Vink and Cassisi (2002), confirming this
scenario. 
Radiative levitation and the ``wind emission'' effect is also the most likely explanation for
the observed jump in the stellar gravities (Moehler et al. 2000, Vink
and Cassisi 2002). 
 
However, the real open problem is that we still lack of any plausible 
explanation for the presence of fast rotators (see discussion in 
R02).

\begin{acknowledgements}
We thank L. R. Bedin, F. Grundahl and A. Rosenberg for gently providing their 
photometries. We are also grateful to S. Lucatello for useful discussions.
We sincerelly thank the anonymous referee for the care and the interest dedicated
to this work and the important contribution to the improvement of this paper.
ARB recognizes the support of the Istituto Nazionale di Astrofisica({\it INAF}).  
GP recognizes partial support from the Ministero dell'Istruzione, Universit\`a e Ricerca ({\it MIUR}), 
and from the Agenzia Spaziale Italiana ({\it ASI}).
\end{acknowledgements}

\begin{table}[h]
\footnotesize
\centering
\begin{tabular}{||ccccccc||}
\hline
\hline
\textbf{Cluster}& \textbf{[Fe/H]} & \textbf{M$_V$} &  \textbf{(m-M)$_V$}& \textbf{E(B-V)}&\textbf{log(r$_t$/r$_c$)} & \textbf{R$_{GC}$ (Kpc)}\\
\hline
\Large{$_{NGC~1904}$}&\Large{$_{-1.37}$}& \Large{$_{-7.86}$}&\Large{$_{15.59}$} & \Large{$_{0.01}$}& \Large{$_{1.72}$}& \Large{$_{18.8  }$} \\
\Large{$_{NGC~2808}$}&\Large{$_{-1.21}$}&\Large{$_{-9.36}$}&\Large{$_{15.56}$}& \Large{$_{0.23}$}& \Large{$_{1.77}$}& \Large{$_{11.0  }$}  \\
\Large{$_{NGC~6093}$}&\Large{$_{-1.43}$}&\Large{$_{-8.23}$}&\Large{$_{15.56}$} & \Large{$_{0.18}$} & \Large{$_{1.95}$} & \Large{$_{~3.8}$}  \\
\Large{$_{NGC~7078}$}&\Large{$_{-2.12}$}&\Large{$_{-9.17}$}&\Large{$_{15.37}$} & \Large{$_{0.10}$} & \Large{$_{2.50c}$} & \Large{$_{10.4 }$}\\
\hline 
\hline                  
\end{tabular}
\normalsize
\caption{The target clusters parameters. [Fe/H] is from Carretta et al.\ 
(2001). The absolute visual magnitude,  distance modulus, central   concentration and
Galactocentric distance are taken from Harris (1999).}
\end{table}
  
\begin{figure}[h]
\centering
\begin{tabular}{l l}
\includegraphics[width= 8 cm]{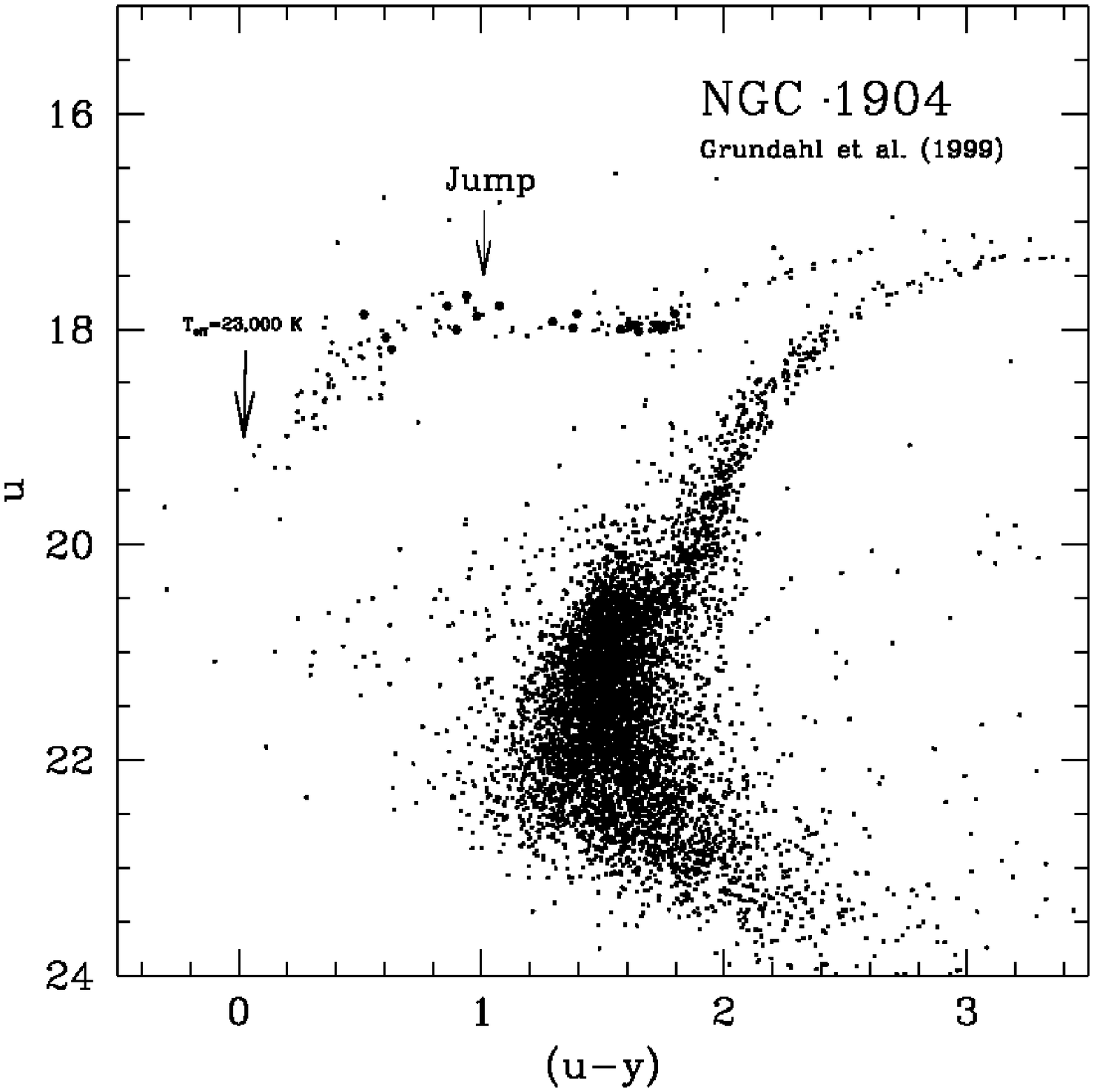} & \includegraphics[width= 8 cm]{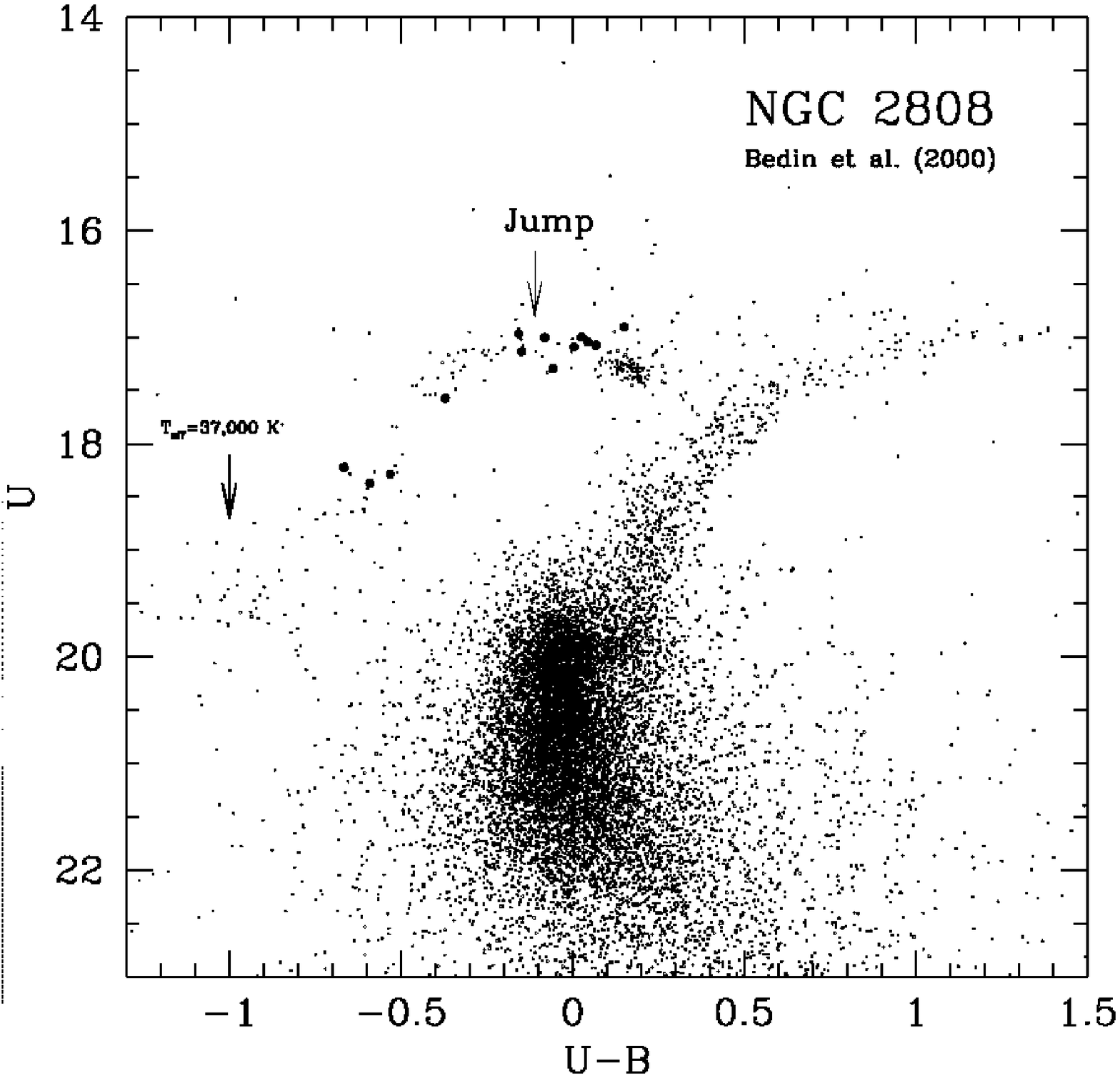}\\
\includegraphics[width= 8 cm]{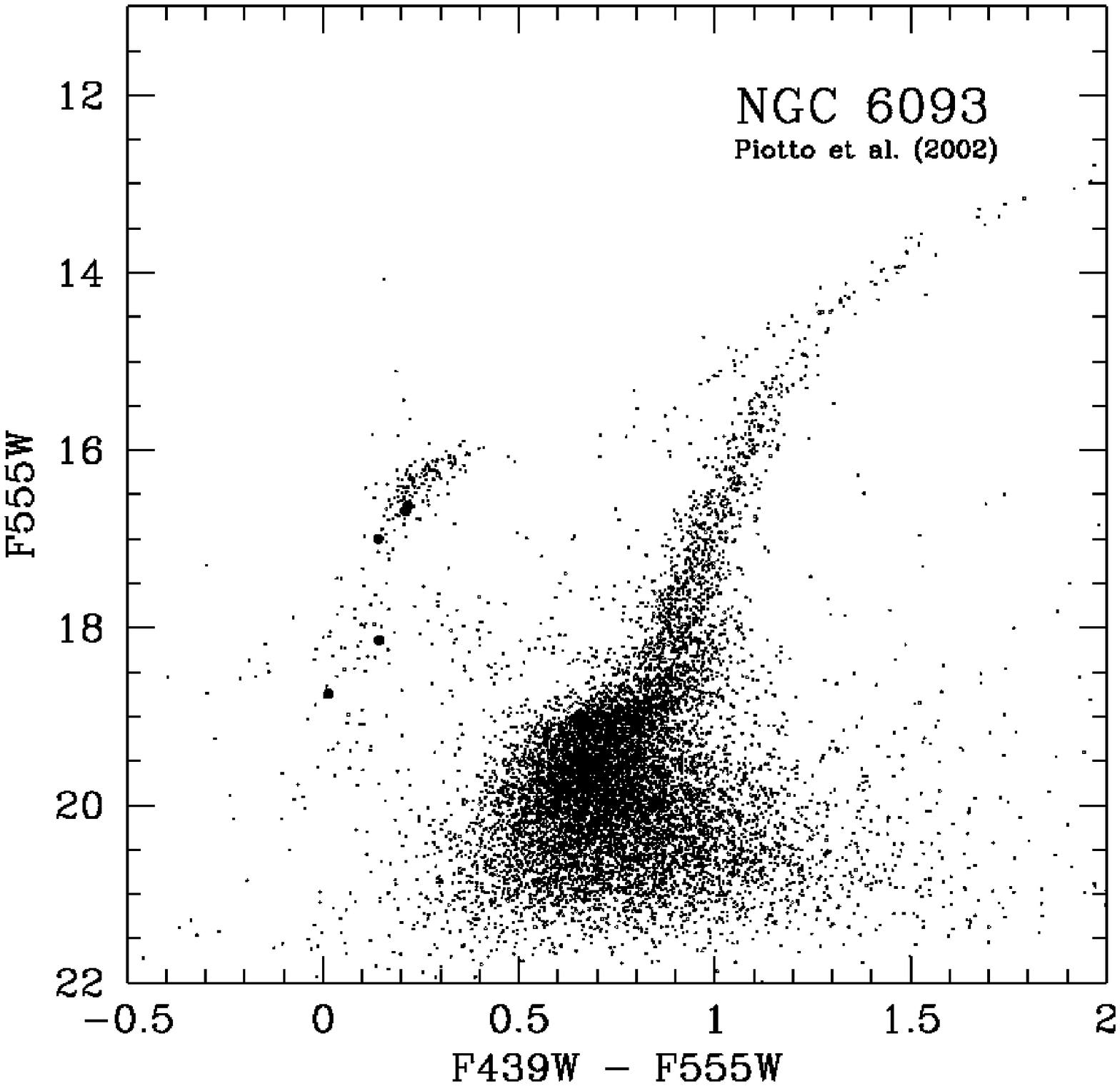} & \includegraphics[width= 8 cm]{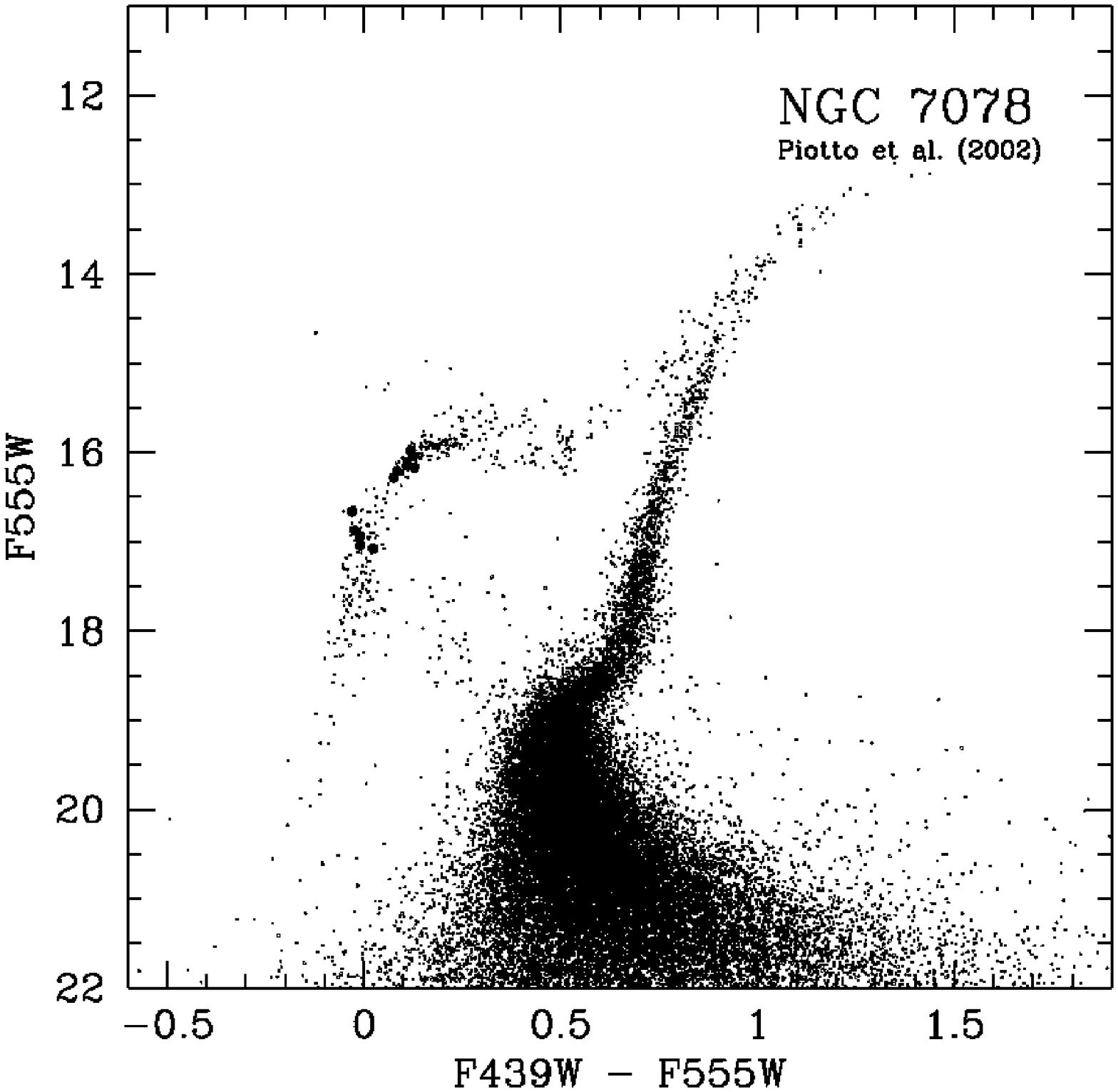}\\
\end{tabular}
\caption{The target cluster CMDs with their corresponding selected stars.}
\end{figure}

\begin{figure*}[t]
\centering
\includegraphics[width=18cm,height=7cm]{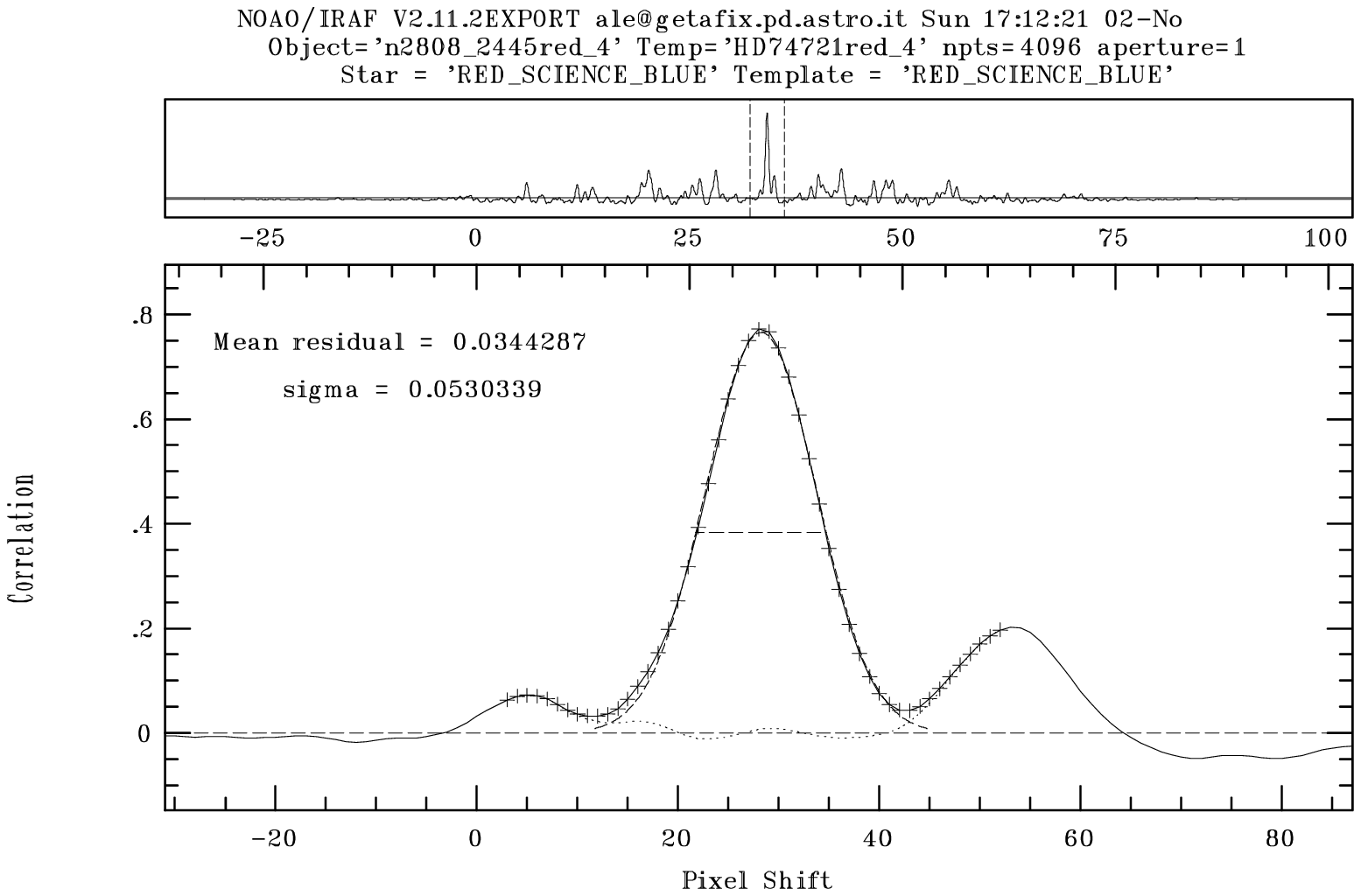}
\includegraphics[width=18cm,height=7cm]{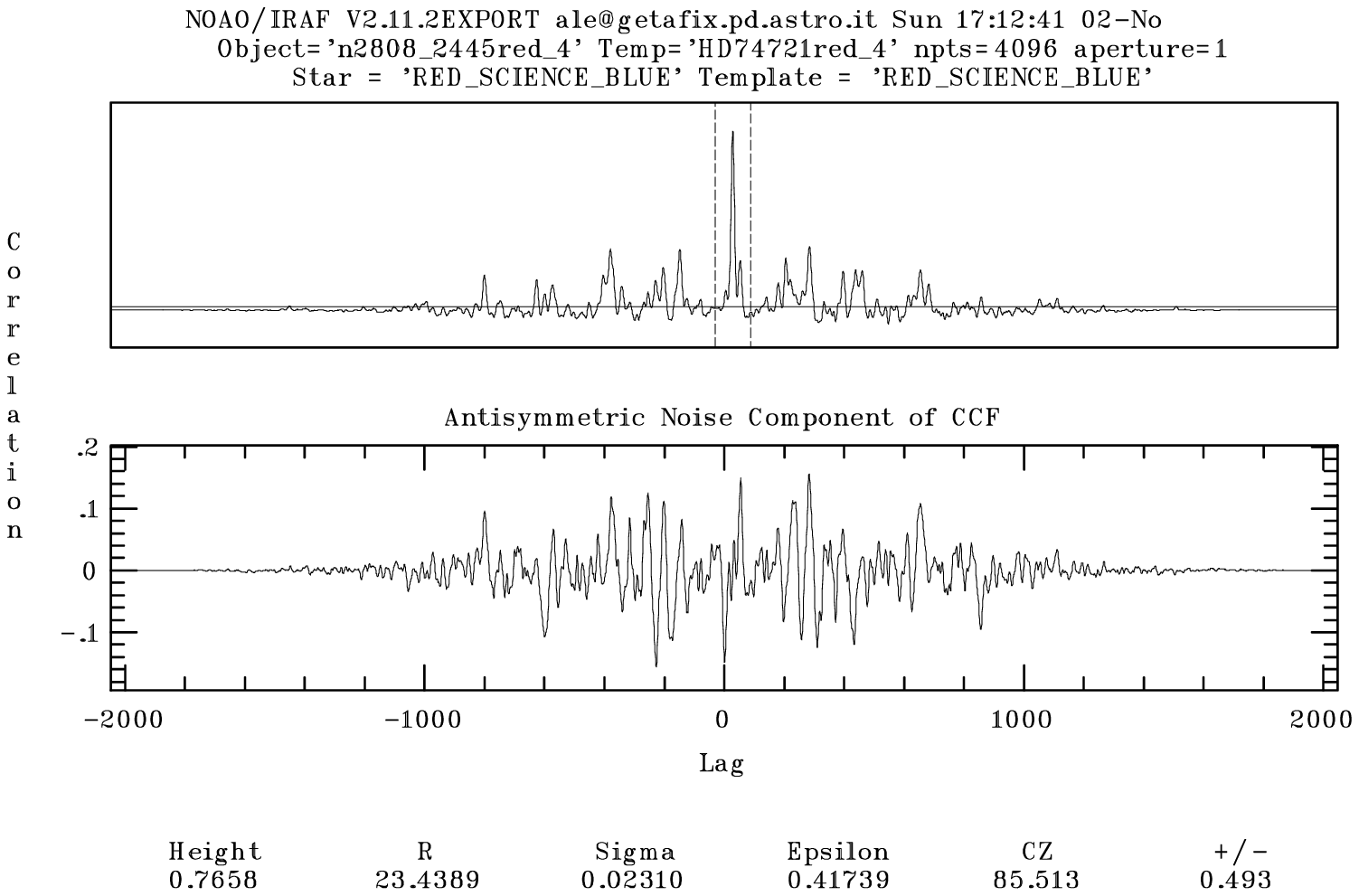}
\caption{Example of a cross-correlation function (top panel), a zoom of the peak (central panel) and the corresponding
antisimmetric part (bottom panel). The target star is n2808-2333 and the template is HD~74721. The values of the r, 
$\epsilon$, $\sigma_{a}$ parameters are also provided. }
\label{f_all6}
\end{figure*}
 
\begin{figure*}[t]
\centering
\includegraphics[width=18cm,height=14cm]{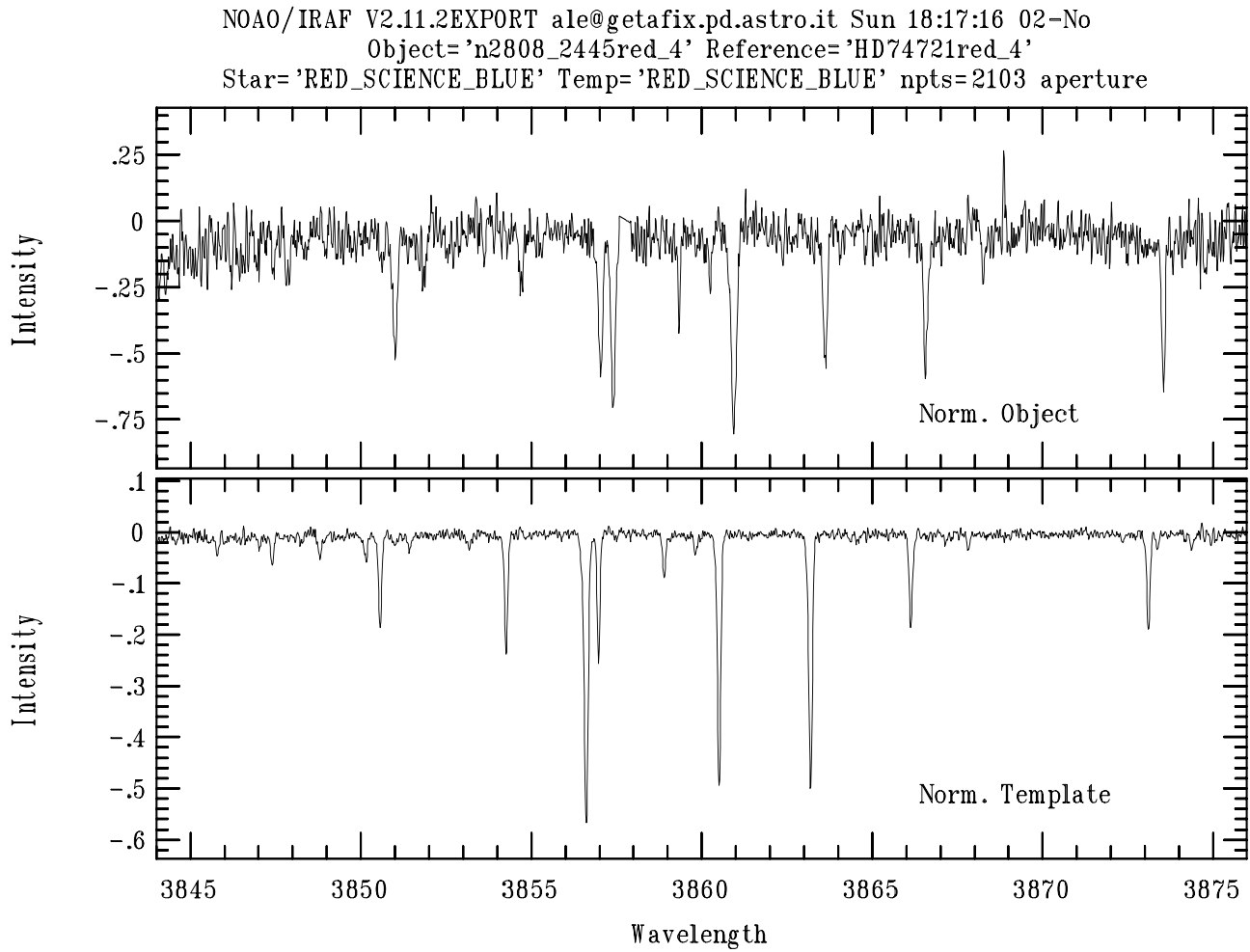}
\caption{Normalized section of the spectrum of the target star n2808-2333 (top panel) and the template 
HD~74721 (bottom panel)}
\label{f_all6}
\end{figure*}
 
\begin{figure*}[t]
\centering
\includegraphics[width=18cm,height=20cm]{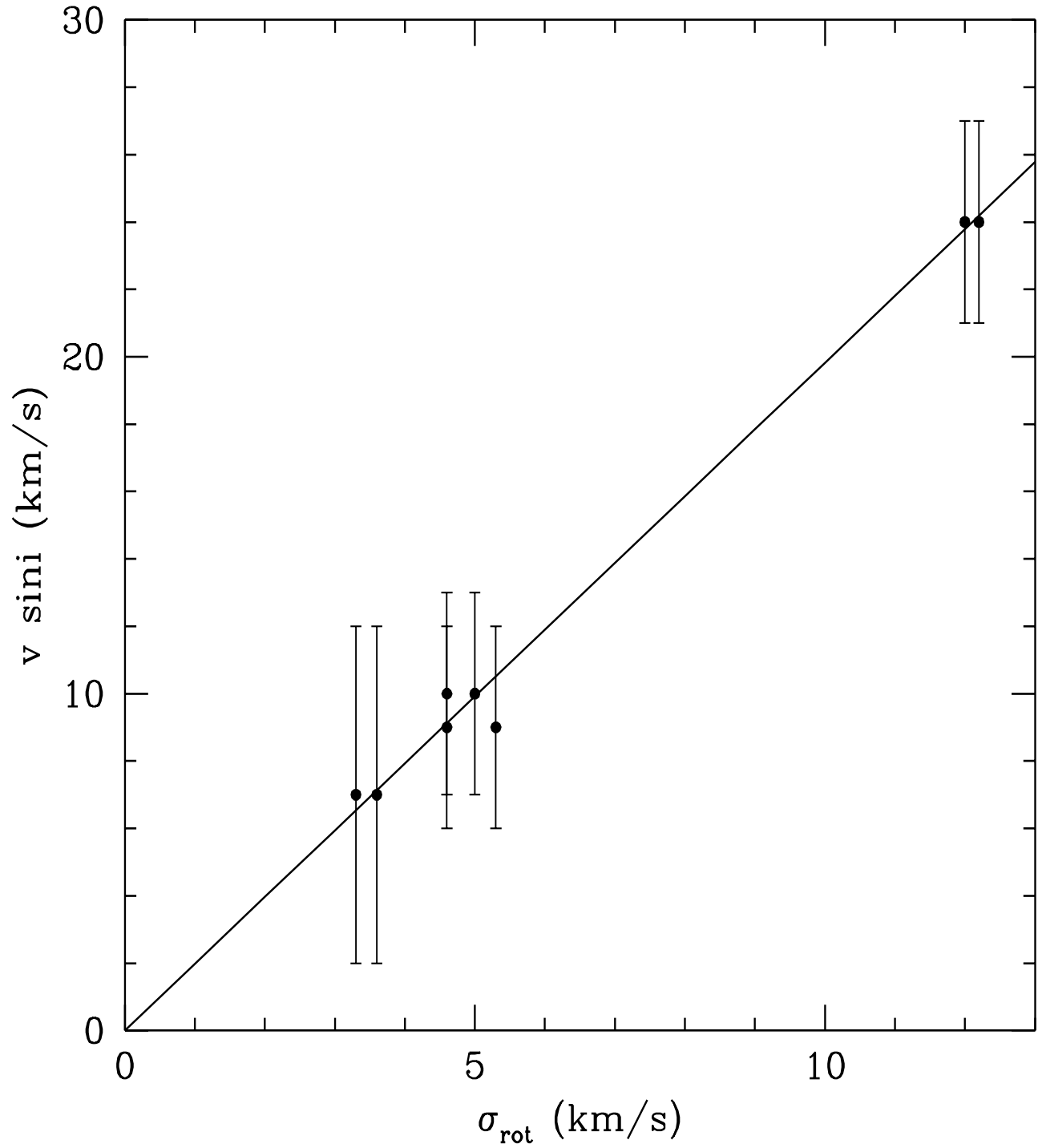}
\caption{Calibration of the differential broadening of the CCF to the projected rotational velocity 
$v$sin$i$ for the standard stars taken from Peterson et al. (1983).}
\label{f_all6}
\end{figure*}

\begin{figure*}[t]
\centering
\includegraphics[width=18cm,height=20cm]{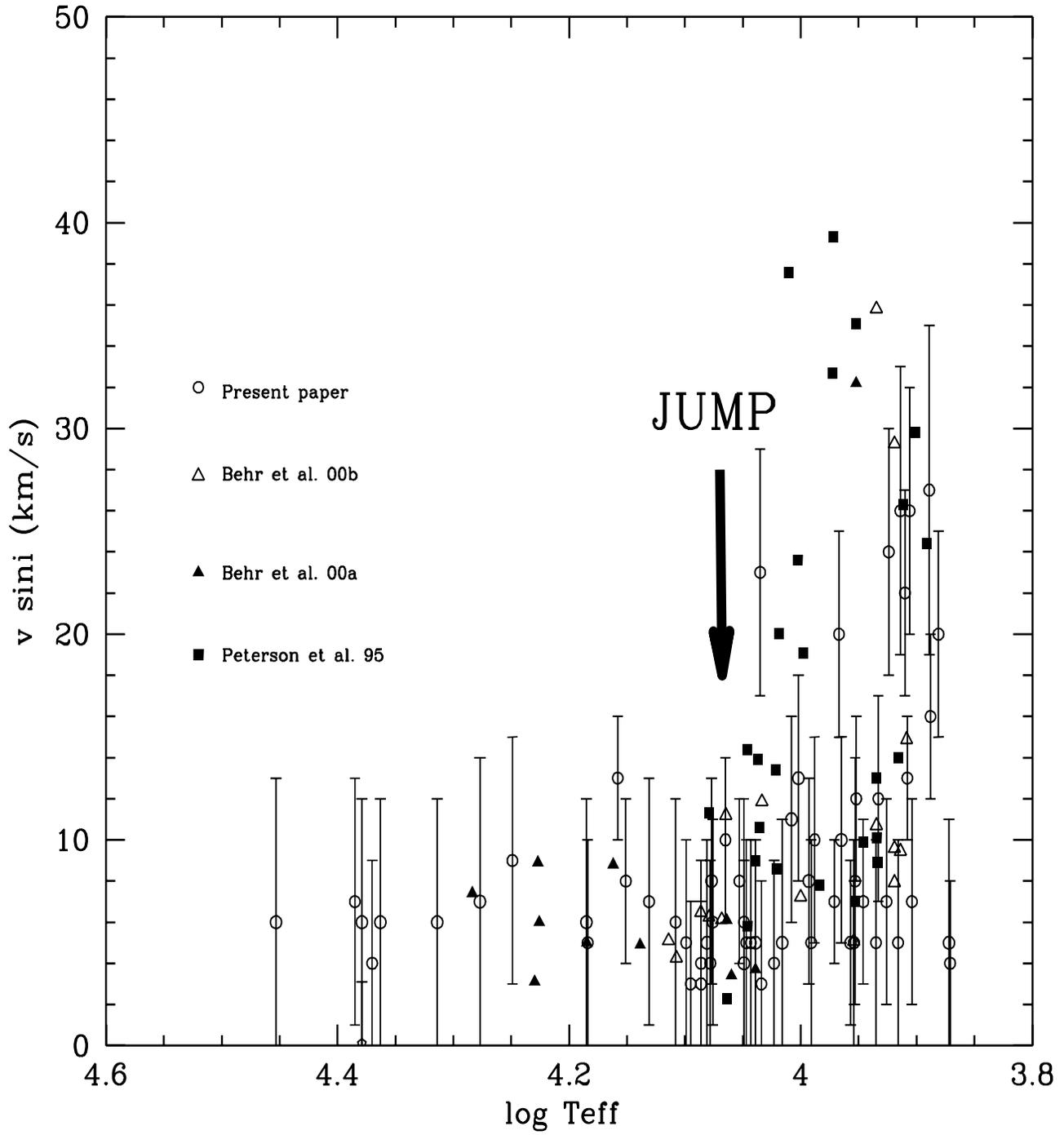}
\caption{Projected rotational 
velocities as a function of the temperature for {\bf all our 61} targets (open circles), plus
the B00b (open triangles), B00a (full triangles)  and {\bf Peterson et al.\ (1995,} full squares) meassurements 
in M15 and M13.  The vertical arrow indicates the 
position of the ``jump'' (from G99) in 
correspondence of $T_{\rm eff}=11,500$K where all the HBs of G99 
GCs show this feature.}
\label{f_all6}
\end{figure*}

\begin{figure*}[t]
\centering
\includegraphics[width=18cm,height=20cm]{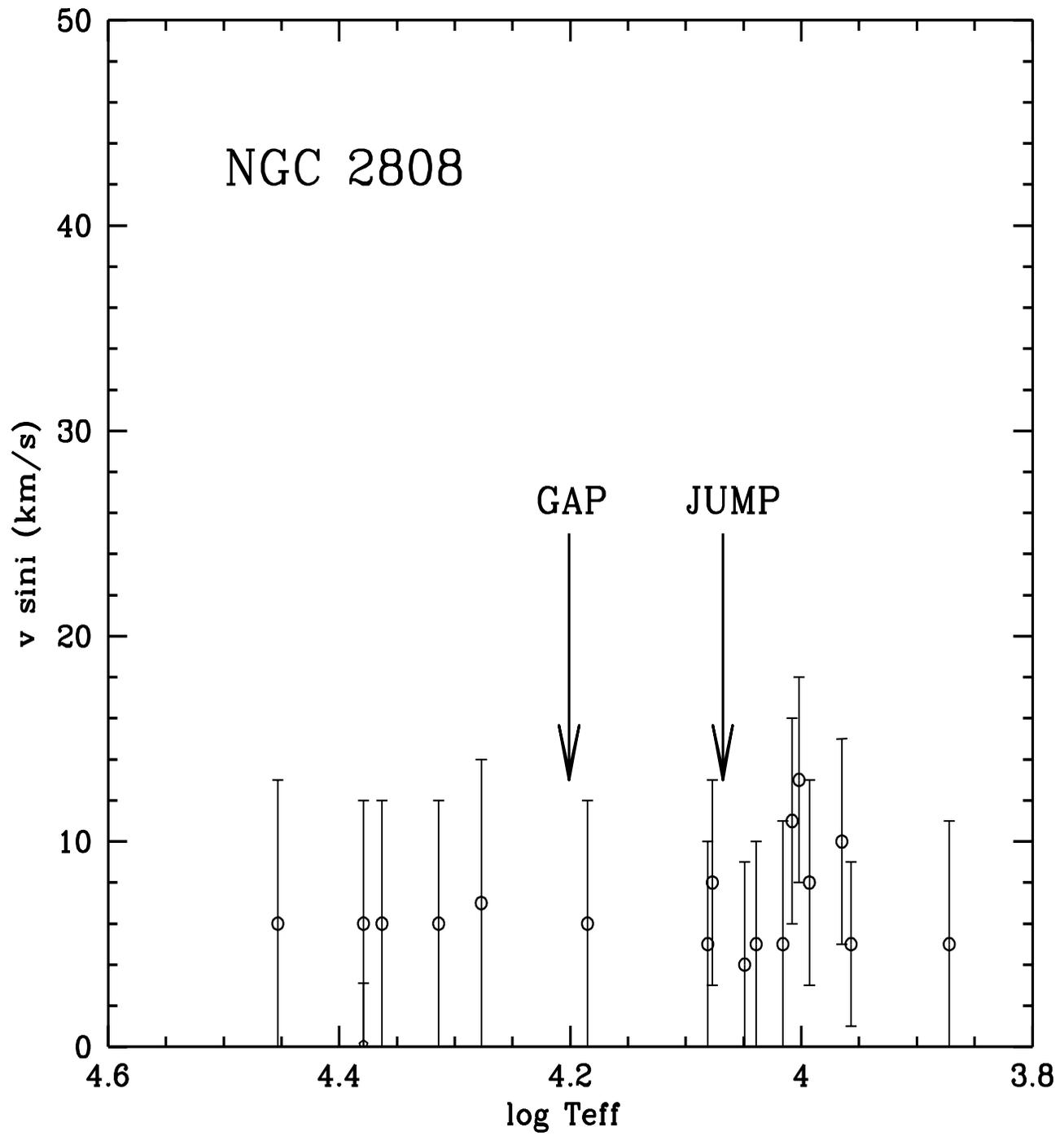}
\caption{Projected rotational velocities as a function of the temperature for NGC~2808.}
\label{f_all6}
\end{figure*}

\centering
\begin{table}[h]
\centering
\begin{tabular}{||l|llllll||} \hline
\hline 
\textbf{Star}&\textbf{$\alpha_{2000}$}&\textbf{$\delta_{2000}$}&\textbf{{\it u}}&\textbf{~~({\it u-y})}&\textbf{S/N}&\textbf{log~(T$_{eff}$)}\\
\hline
 m79-489 & 05:24:10.6 & $-$24:32:46 &   17.861  &  ~~0.516 & 11 &  4.184 \\
 m79-535 & 05:24:16.6 & $-$24:31:59 &   18.075  &  ~~0.607 & 15 &  4.158 \\ 
 m79-555 & 05:24:15.8 & $-$24:30:36 &   18.183  &  ~~0.631 & 14 &  4.151 \\ 
 m79-434 & 05:24:16.6 & $-$24:31:37 &   17.782  &  ~~0.859 & 10 &  4.095 \\ 
 m79-469 & 05:24:08.2 & $-$24:32:13 &   18.004  &  ~~0.897 & 11 &  4.086 \\ 
 m79-392 & 05:24:13.4 & $-$24:30:25 &   17.684  &  ~~0.939 & 30 &  4.076 \\ 
 m79-363 & 05:24:13.4 & $-$24:31:26 &   17.879  &  ~~0.982 & 15 &  4.065 \\ 
 m79-389 & 05:24:16.3 & $-$24:31:30 &   17.781  &  ~~1.075 & 13 &  4.043 \\
 m79-354 & 05:24:07.9 & $-$24:32:17 &   17.929  &  ~~1.294 & 10 &  3.991 \\ 
 m79-366 & 05:24:15.4 & $-$24:31:26 &   17.987  &  ~~1.378 & 14 &  3.971 \\ 
 m79-297 & 05:24:17.2 & $-$24:30:18 &   17.852  &  ~~1.394 & ~~9&  3.967 \\ 
 m79-327 & 05:24:14.0 & $-$24:29:54 &   18.000  &  ~~1.574 & 10 &  3.924 \\ 
 m79-281 & 05:24:27.2 & $-$24:29:16 &   17.958  &  ~~1.609 & ~~8&  3.916 \\ 
 m79-298 & 05:24:21.6 & $-$24:28:59 &   17.967  &  ~~1.615 & 10 &  3.914 \\ 
 m79-289 & 05:24:17.5 & $-$24:31:23 &   17.957  &  ~~1.635 & ~~9&  3.910 \\ 
 m79-295 & 05:24:02.0 & $-$24:31:35 &   18.016  &  ~~1.648 & 15 &  3.906 \\ 
 m79-275 & 05:24:17.9 & $-$24:30:40 &   17.970  &  ~~1.723 & ~~8&  3.889 \\ 
 m79-243 & 05:24:19.6 & $-$24:32:26 &   17.959  &  ~~1.725 & 13 &  3.888 \\ 
 m79-294 & 05:24:18.9 & $-$24:31:49 &   17.999  &  ~~1.754 & 10 &  3.881 \\ 
 m79-209 & 05:24:16.2 & $-$24:32:16 &   17.852  &  ~~1.797 & 16 &  3.871 \\ 
\hline
\hline
\hline
\textbf{Star}&\textbf{$\alpha_{2000}$}&\textbf{$\delta_{2000}$}&\textbf{{\it U}}&\textbf{~~({\it U-B})}&\textbf{S/N} &\textbf{log~(T$_{eff}$)}\\
\hline
 n2808-7596  & 9:11:24.9 & $-$64:52:46 &  17.714 &  $-$0.785   & ~~7&   4.453 \\
 n2808-9432  & 9:12:23.8 & $-$64:52:08 &  18.219 &  $-$0.666   & ~~8&   4.363 \\
 n2808-11222 & 9:12:53.6 & $-$64:51:05 &  18.370 &  $-$0.591   & ~~7&   4.314 \\      
 n2808-9655  & 9:11:59.8 & $-$64:47:45 &  18.287 &  $-$0.532   & ~~9&   4.277 \\  
 n2808-6427  & 9:11:41.6 & $-$64:49:16 &  17.574 &  $-$0.371   & ~~7&   4.185 \\  
 n2808-3159  & 9:12:12.2 & $-$64:58:23 &  16.967 &  $-$0.157   & ~~7&   4.081 \\     
 n2808-4512  & 9:11:18.7 & $-$64:57:14 &  17.136 &  $-$0.149   & ~~9&   4.077 \\  
 n2808-3715  & 9:12:25.9 & $-$64:55:13 &  17.003 &  $-$0.081   & ~~8&   4.049 \\   
 n2808-4991  & 9:11:24.2 & $-$64:49:59 &  17.294 &  $-$0.057   & 11 &   4.039 \\   
 n2808-3721  & 9:11:58.1 & $-$64:57:09 &  17.092 &   ~~0.004   & 10 &   4.016 \\    
 n2808-3435  & 9:12:18.3 & $-$64:49:37 &  16.999 &   ~~0.026   & 15 &  4.008  \\
 n2808-3841  & 9:11:49.8 & $-$64:57:30 &  17.043 &   ~~0.044   & 20 &   4.002 \\
 n2808-3949  & 9:12:17.2 & $-$64:46:41 &  17.077 &   ~~0.068   & ~~7&   3.993 \\
 n2808-2333  & 9:12:23.4 & $-$64:52:00 &  16.804 &   ~~0.174   & 22 &   3.957 \\
 n2808-2445  & 9:11:42.3 & $-$64:52:51 &  16.853 &   ~~0.290   & 25 &  3.872  \\
 n2808-2909  & 9:12:03.9 & $-$64:49:27 &  16.904 &   ~~0.150   & 20 &  3.965  \\  
\hline	     
\hline	     
\end{tabular}
\caption{Positions, photometric information and effective temperature  of target stars in M79 and NGC~2808.}
\end{table}

\centering
\begin{table}[h]
\centering
\begin{tabular}{||l|llllll||} \hline
\hline 
\textbf{Star}&\textbf{$\alpha_{2000}$}&\textbf{$\delta_{2000}$}&\textbf{{\it F555W}}&\textbf{~~({\it F439W-F555W})}&\textbf{S/N}
&\textbf{log~(T$_{eff}$)}\\
\hline
m80-509   & 16:17:05.2 & $-$22:58:47 &  16.622 &  ~~0.216    &  12 &  3.946  \\
 m80-820   & 16:17:08.2 & $-$22:58:30 &  16.684 &  ~~0.209   &  14 &  3.953 \\
 m80-1400  & 16:17:08.5 & $-$22:59:00 &  18.141 &  ~~0.143   &  10 &  4.047  \\
 m80-2044  & 16:17:05.8 & $-$22:57:54 &  16.998 &  ~~0.141   &  12 &  4.053 \\
 m80-2242  & 16:17:07.2 & $-$22:57:56 &  18.740 &  ~~0.014   &  10 &  4.385 \\
 m80-938   & 16:17:08.2 & $-$22:58:30 &  16.886 & $-$0.150   &  11 &  4.099 \\	 
\hline 
&  & &\textbf{{\it V}}&\textbf{~~({\it V-I})}& &\\
\hline
 m80-107   & 16:17:02.5 & $-$23:02:25 &  16.88  &   ~~0.221  &  10 &  4.049  \\  
 m80-109   & 16:17:00.1 & $-$23:00:44 &  16.86  &   ~~0.232  &  10 &  4.034  \\
 m80-83    & 16:17:17.9 & $-$22:58:27 &  16.57  &   ~~0.270  &  11 &  3.988  \\
 m80-1149  & 16:17:03.3 & $-$23:00:17 &  18.15  &   ~~0.095  &  10 &  4.249  \\  
 m80-454   & 16:17:11.8 & $-$22:58:05 &  18.70  &   ~~0.034  &  10 &  4.370  \\
\hline
\hline
\hline
\textbf{Star}&\textbf{$\alpha_{2000}$}&\textbf{$\delta_{2000}$}&\textbf{{\it V}}&\textbf{~~({\it B-V})}&\textbf{S/N}
&\textbf{log~(T$_{eff}$)}\\
\hline
 m15-1813  & 21:30:03.8 & 12:10:26 & 16.04 & ~~0.124 &  16 & 3.930 \\
 m15-5516  & 21:30:01.0 & 12:11:18 & 16.28 & ~~0.076 &  17 & 3.952 \\
 m15-6143  & 21:29:59.1 & 12:08:37 & 16.21 & ~~0.084 &  18 & 3.954 \\ 
 m15-2700  & 21:30:00.1 & 12:10:34 & 16.66 &$-$0.031 &  13 & 4.131 \\ 
 m15-2917  & 21:29:57.9 & 12:11:08 & 15.98 & ~~0.119 &  15 & 3.926 \\
 m15-3333  & 21:30:04.0 & 12:09:49 & 16.17 & ~~0.127 &  16 & 3.933 \\  
 m15-4047  & 21:29:59.0 & 12:11:10 & 16.15 & ~~0.110 &  16 & 3.935 \\    
 m15-4536  & 21:30:00.0 & 12:11:05 & 17.08 & ~~0.023 &  15 & 4.023  \\ 
 m15-5168  & 21:30:00.0 & 12:08:45 & 16.88 &$-$0.022 &  15 & 4.108  \\ 
 m15-768   & 21:30:03.8 & 12:10:46 & 16.94 &$-$0.011 &  14 & 4.086 \\
 m15-817   & 21:30:02.8 & 12:10:30 & 17.04 &$-$0.010 &  16 & 4.078  \\
 m15-b130  & 21:29:54.8 & 12:12:15 & 15.96 & ~~0.150 &  15 & 3.954  \\
 m15-b218  & 21:29:45.6 & 12:11:25 & 15.99 & ~~0.160 &  17 & 3.908  \\
\hline 
&  & &\textbf{{\it V}}&\textbf{~~({\it V-I})}& &\\
\hline
 m15-1048  & 21:30:03.3 & 12:12:02 & 16.79 & ~~0.027 & 16 & 4.058 \\
 m15-699   & 21:29:47.0 & 12:09:51 & 16.63 & ~~0.064 & 14 & 4.035 \\
 m15-788   & 21:29:49.5 & 12:12:37 & 15.94 & ~~0.239 & 15 & 3.904 \\
\hline	     
\hline	     
\end{tabular}
\caption{Positions, photometric information and effective temperature and of target stars
in M80 and M15.} 
\end{table}  

\centering
\begin{table}[h]
\footnotesize
\centering
\begin{tabular}{||l|llllll||}
\hline 
\textbf{Star}&\textbf{$v_{rad}$}&\textbf{$v$sin$i$}& $|\Delta(v$sin$i)|_{mes}$& $|\Delta(v$sin$i)|_{t}$ &$|\Delta(v$sin$i)|_{A}$ & $|\Delta(v$sin$i)|_{tot}$\\
& (km/s) & (km/s) & (km/s) & (km/s) & (km/s) & (km/s)\\
\hline
 m79-489 & $+$196.1~$\pm$~0.9 &  ~~5 & $\pm$4.7 & $\pm$2.3& $\pm$0.3& $\pm$5\\
 m79-535 & $+$199.2~$\pm$~0.8 &   13 & $\pm$2.8 & $\pm$0.3& $\pm$1.7& $\pm$3\\ 
 m79-555 & $+$201.5~$\pm$~0.9 &  ~~8 & $\pm$4.0 & $\pm$1.3& $\pm$0.8& $\pm$4\\ 
 m79-434 & $+$207.3~$\pm$~1.0 &  ~~3 & $\pm$2.5 & $\pm$3.0& $\pm$0.0& $\pm$4\\ 
 m79-469 & $+$202.6~$\pm$~0.8 &  ~~3 & $\pm$2.6 & $\pm$3.0& $\pm$0.0& $\pm$4\\ 
 m79-392 & $+$204.3~$\pm$~0.8 &  ~~6 & $\pm$4.8 & $\pm$1.9& $\pm$0.5& $\pm$5\\ 
 m79-363 & $+$198.7~$\pm$~0.8 &   10 & $\pm$3.7 & $\pm$0.7& $\pm$1.2& $\pm$4\\ 
 m79-389 & $+$195.8~$\pm$~0.9 &  ~~5 & $\pm$4.5 & $\pm$2.3& $\pm$0.3& $\pm$5\\
 m79-354 & $+$209.0~$\pm$~1.3 &  ~~5 & $\pm$4.5 & $\pm$2.3& $\pm$0.3& $\pm$5\\ 
 m79-366 & $+$205.1~$\pm$~1.1 &  ~~7 & $\pm$2.4 & $\pm$1.7& $\pm$0.7& $\pm$3\\ 
 m79-297 & $+$202.2~$\pm$~1.7 &   20 & $\pm$2.4 & $\pm$2.6& $\pm$2.8& $\pm$5\\ 
 m79-327 & $+$204.2~$\pm$~1.6 &   24 & $\pm$3.6 & $\pm$4.0& $\pm$3.5& $\pm$6\\ 
 m79-281 & $+$213.1~$\pm$~1.8 &  ~~5 & $\pm$4.5 & $\pm$2.3& $\pm$0.3& $\pm$5\\ 
 m79-298 & $+$214.2~$\pm$~2.2 &   26 & $\pm$3.6 & $\pm$4.6& $\pm$3.8& $\pm$7\\ 
 m79-289 & $+$205.5~$\pm$~1.8 &   22 & $\pm$1.8 & $\pm$3.3& $\pm$3.2& $\pm$5\\ 
 m79-295 & $+$208.2~$\pm$~1.1 &   26 & $\pm$1.8 & $\pm$4.6& $\pm$3.8& $\pm$6\\ 
 m79-275 & $+$206.5~$\pm$~2.9 &   27 & $\pm$5.6 & $\pm$4.9& $\pm$3.9& $\pm$8\\ 
 m79-243 & $+$207.1~$\pm$~0.6 &   16 & $\pm$2.8 & $\pm$1.3& $\pm$2.2& $\pm$4\\ 
 m79-294 & $+$208.9~$\pm$~1.5 &   20 & $\pm$2.6 & $\pm$2.6& $\pm$2.8& $\pm$5\\ 
 m79-209 & $+$199.1~$\pm$~0.4 &  ~~3 & $\pm$2.7 & $\pm$3.0& $\pm$0.0& $\pm$4\\ 
\hline
\hline
 n2808-7596  & $+$101.6~$\pm$~5.4 &  ~~6 & $\pm$6.3 &$\pm$1.8&$\pm$0.6& $\pm$7\\
 n2808-9432  & $+$104.5~$\pm$~3.5 &  ~~6 & $\pm$5.7 &$\pm$2.0&$\pm$0.5& $\pm$6\\
 n2808-11222 & $+$105.7~$\pm$~2.6 &  ~~6 & $\pm$5.6 &$\pm$2.0&$\pm$0.5& $\pm$6\\      
 n2808-9655  & $+$106.3~$\pm$~1.7 &  ~~7 & $\pm$6.8 &$\pm$1.7&$\pm$0.7& $\pm$7\\  
 n2808-6427  & $+$102.7~$\pm$~1.4 &  ~~6 & $\pm$5.7 &$\pm$2.0&$\pm$0.5& $\pm$6\\  
 n2808-3159  & $+$102.4~$\pm$~1.2 &  ~~5 & $\pm$4.8 &$\pm$2.3&$\pm$0.3& $\pm$5\\     
 n2808-4512  & $+$101.1~$\pm$~1.0 &  ~~8 & $\pm$4.7 &$\pm$1.3&$\pm$0.8& $\pm$5\\  
 n2808-3715  & $+$101.3~$\pm$~1.2 &  ~~4 & $\pm$4.4 &$\pm$2.5&$\pm$0.2& $\pm$5\\   
 n2808-4991  & $+$102.9~$\pm$~1.2 &  ~~5 & $\pm$4.8 &$\pm$2.3&$\pm$0.3& $\pm$5\\   
 n2808-3721  & $+$101.6~$\pm$~1.2 &  ~~6 & $\pm$5.7 &$\pm$2.0&$\pm$0.5& $\pm$6\\    
 n2808-3435  & $+$114.5~$\pm$~1.4 &   11 & $\pm$4.5 &$\pm$0.3&$\pm$1.3& $\pm$5\\
 n2808-3841  & ~~$+$96.4~$\pm$~1.2&   13 & $\pm$3.4 &$\pm$0.3&$\pm$1.7& $\pm$4\\
 n2808-3949  & $+$103.3~$\pm$~1.3 &  ~~8 & $\pm$5.6 &$\pm$1.3&$\pm$0.8& $\pm$6\\
 n2808-2333  & $+$109.4~$\pm$~0.6 &  ~~5 & $\pm$4.8 &$\pm$2.3&$\pm$0.3& $\pm$5\\
 n2808-2445  & ~~$+$85.7~$\pm$~0.4&  ~~5 & $\pm$4.7 &$\pm$2.3&$\pm$0.3& $\pm$5\\
 n2808-2909  & $+$105.6~$\pm$~1.2 &   10 & $\pm$4.7 &$\pm$0.7&$\pm$1.2& $\pm$5\\ 
\hline	     
\hline	     
\end{tabular}
\caption{Velocity measurements of target stars in M79 and NGC~2808.}
\end{table}

\centering
\begin{table}[h]
\footnotesize
\centering
\begin{tabular}{||l|llllll||}
\hline 
\textbf{Star}&\textbf{$v_{rad}$}&\textbf{$v$sin$i$}& $|\Delta(v$sin$i)|_{mes}$& $|\Delta(v$sin$i)|_{t}$&$|\Delta(v$sin$i)|_{A}$& $|\Delta(v$sin$i)|_{tot}$\\
& (km/s) & (km/s) & (km/s) & (km/s) & (km/s) & (km/s)\\
\hline
 m80-509   & ~~~$+$7.9~$\pm$~0.5 & ~~7 & $\pm$3.7&$\pm$1.7& $\pm$ 0.7& $\pm$4\\
 m80-820   & ~~~$+$8.6~$\pm$~1.1 & ~~8 & $\pm$5.8&$\pm$1.3& $\pm$ 0.8& $\pm$6\\
 m80-1400  & ~~~$+$5.7~$\pm$~0.7 & ~~5 & $\pm$4.7&$\pm$2.3& $\pm$ 0.3& $\pm$5\\
 m80-2044  & ~~$+$23.5~$\pm$~0.5 & ~~8 & $\pm$3.6&$\pm$1.3& $\pm$ 0.8& $\pm$4\\
 m80-2242  & ~~$+$26.6~$\pm$~1.2 & ~~7 & $\pm$5.6&$\pm$1.7& $\pm$ 0.7& $\pm$6\\
 m80-938   & ~~$+$17.8~$\pm$~0.8 & ~~5 & $\pm$4.7&$\pm$2.3& $\pm$ 0.3& $\pm$5\\
 m80-107   & ~~$+$15.9~$\pm$~0.9 & ~~6 & $\pm$5.6&$\pm$2.0& $\pm$ 0.5& $\pm$6\\  
 m80-109   & ~~$+$14.1~$\pm$~0.6 & ~~3 & $\pm$3.6&$\pm$3.0& $\pm$ 0.0& $\pm$5\\
 m80-83    & ~~$+$10.0~$\pm$~1.1 &  10 & $\pm$4.5&$\pm$0.7& $\pm$ 1.2& $\pm$5\\
 m80-1149  & ~~$+$18.3~$\pm$~1.4 & ~~9 & $\pm$5.7&$\pm$1.0& $\pm$ 1.0& $\pm$6\\  
 m80-454   & ~~~~~$-$0.7~$\pm$~1.0 & ~~4 &$\pm$4.2&$\pm$2.5&$\pm$ 0.2& $\pm$5\\
\hline
\hline
 m15-1813  & ~~$-$98.7~$\pm$~1.0&  10 &$\pm$4.5&$\pm$0.7&$\pm$ 1.2& $\pm$5\\
 m15-5516  & ~~$-$99.0~$\pm$~1.2&  12 &$\pm$3.6&$\pm$0.0&$\pm$ 1.5& $\pm$4\\
 m15-6143  & $-$101.7~$\pm$~0.8 & ~~5 &$\pm$4.5&$\pm$2.3&$\pm$ 0.3& $\pm$5\\ 
 m15-2700  & $-$148.0~$\pm$~0.9 & ~~7 &$\pm$5.8&$\pm$1.7&$\pm$ 0.7& $\pm$6\\ 
 m15-2917  & $-$109.7~$\pm$~0.7 & ~~7~&$\pm$4.8&$\pm$1.7&$\pm$ 0.7& $\pm$5\\
 m15-3333  & ~~$-$91.9~$\pm$~1.6&  12 &$\pm$4.5&$\pm$0.3&$\pm$ 1.5& $\pm$5\\  
 m15-4047  & ~~$-$93.1~$\pm$~1.5& ~~5 &$\pm$4.7&$\pm$2.3&$\pm$ 0.3& $\pm$5\\    
 m15-4536  & $-$101.9~$\pm$~1.2 & ~~4 &$\pm$4.2&$\pm$2.5&$\pm$ 0.2& $\pm$5\\ 
 m15-5168  & $-$102.9~$\pm$~0.9 & ~~6 &$\pm$5.7&$\pm$2.0&$\pm$ 0.5& $\pm$6\\ 
 m15-768   & $-$102.6~$\pm$~0.9 & ~~4 &$\pm$3.8&$\pm$2.7&$\pm$ 0.2& $\pm$5\\
 m15-817   & $-$110.8~$\pm$~1.3 & ~~4 &$\pm$4.3&$\pm$2.7&$\pm$ 0.2& $\pm$5\\
 m15-b130  & $-$113.9~$\pm$~0.6 & ~~5 &$\pm$2.5&$\pm$2.3&$\pm$ 0.3& $\pm$3\\
 m15-b218  & ~~$-$94.6~$\pm$~0.5&  13 &$\pm$2.1&$\pm$0.3&$\pm$ 1.7& $\pm$3\\
 m15-1048  & $-$101.2~$\pm$~0.6 &  11 &$\pm$3.7&$\pm$0.3&$\pm$ 1.3& $\pm$4\\
 m15-699   & $-$103.2~$\pm$~2.7 &  23 &$\pm$3.4&$\pm$3.7&$\pm$ 3.3& $\pm$6\\
 m15-788   & $-$103.0~$\pm$~0.7 & ~~7 &$\pm$4.8&$\pm$1.7&$\pm$ 0.7& $\pm$5\\
\hline	     
\hline	     
\end{tabular}
\caption{Velocity measurements of target stars in M80 and M15.} 
\end{table}

\end{document}